\documentclass [12pt]{article}
\usepackage{amsmath, amsthm, amssymb}
\usepackage{graphicx}
\usepackage{mathptmx}
\usepackage{fullpage}
\usepackage{authblk}
\usepackage{mathrsfs}
\usepackage{makecell}
\usepackage{bibentry, natbib}
\usepackage{float}
\usepackage{tikz}
\usepackage{pgflibraryarrows}
\usepackage{pgflibrarysnakes}

\newtheorem{Theorem}{Theorem}

\newtheorem{Definition}{Definition}
\newtheorem{Proposition}{Proposition}
\theoremstyle{definition}
\newtheorem{example}{Example}

\title{The probabilistic rank random assignment rule and its axiomatic characterization\footnote{This paper is based on Chapter 7 of Chen's doctoral thesis submitted to Waseda University library in 2013 and combines part of the results of \cite{2018harless}. The authors thank Tsuyoshi Adachi, Xiang Han, Peng Liu, Takashi Oginuma, Koichi Suga, Qianfeng Tang, William Thomson, Jun Wako, Huaxia Zeng, Yongchao Zhang and seminar participants at SUFE, 2012 Japanese Economic Association Autumn meeting, and ISDG 2017 for their helpful comments. This research is supported by The National Natural Science Foundation of China (No. 71703038), The Ministry of Education Project of Youth Fund of Humanities and Social Sciences (No. 17YJC790012), and The Major Project of Humanities and Social Sciences of Shanghai Municipal Education Commission Science and Innovation Plan Program (No. 2017-01-07-00-02-E00008).}}

\author{Yajing Chen\footnote{School of Business, East China University of Science and Technology, 130 Meilong Road, Shanghai 200237, China. Email: yajingchen@ecust.edu.cn, yajingchen87@gmail.com}~~~~Patrick Harless\footnote{University of Arizona, Tucson, United States. Email: pharless@email.arizona.edu.}~~~~Zhenhua Jiao\footnote{School of Business, Shanghai University of International Business and Economics, Shanghai, 201620, China. Email: jzhenhua@163.com.} }
\begin{document}

\maketitle
\begin{abstract}
This paper considers the problem of randomly assigning a set of objects to a set of agents based on the ordinal preferences of agents. We generalize the well-known immediate acceptance algorithm to the afore-mentioned random environments and define the \emph{probabilistic rank rule}\emph{ (PR rule)}. We introduce two new axioms: \emph{sd-rank-fairness},\footnote{Throughout this paper, ``sd" is the short form of  ``stochastic dominance".} and \emph{equal-rank envy-freeness}. Sd-rank-fairness implies sd-efficiency.\footnote{\cite{bogomolnaia2001new} call this concept  ``ordinal efficiency".} Equal-rank envy-freeness implies equal treatment of equals. Sd-rank-fairness and equal-rank envy-freeness are enough to characterize the PR rule.
\newline
\newline
\textbf{\emph{Keywords}}: Random assignment; Probabilistic rank rule; Sd-rank-fairness; Equal-rank envy-freeness; Characterization
\newline
\newline
\textbf{\emph{JEL classification}}: C70; C78; D71; D78
\end{abstract}

\section{Introduction}

In the school choice setting, the Boston mechanism, building on the immediate acceptance algorithm, has been studied extensively along various directions since the seminal research of \cite{abdulkadiroglu2003school}. Given a set of students, a set of schools, the quota vector, the preference profile of students over schools, and the priority profile of schools over  students, the \textbf{immediate acceptance algorithm} works in the following procedure:

\emph{Step $1$}, each student applies to her first-choice school. If the number of students applying to a school, say $a$, is less than its quota, say $q_a$, then all the applicants are permanently accepted by this school. If the number of students applying to school $a$ is greater than its quota, then school $a$  accepts students with the highest $q_a$ priority and reject the other applicants. The quota of one school reduces by the number of students it accepted in this step.

\emph{Step $t(\geqslant2)$}, each student who is rejected at Step $t-1$ applies to her $t^{th}$-choice school. If the number of students applying to a school, say $a$, is less than its remaining quota, say $q_a^t$, then all the applicants are permanently accepted by this school. If the number of students applying to school $a$ is greater than its remaining quota, then school $a$  accepts students with the highest $q_a^t$ priority and reject the other applicants. The quota of one school reduces again by the number of students it accepted in this step.

The algorithm terminates when no students are rejected.

Apart from school choice, this paper considers the problem of randomly assigning a set of indivisible objects to a set of agents based on ordinal preferences of agents without monetary transfers (\cite{shapley1974cores} and \cite{hylland1979efficient}). This problem has a number of applications, such as dormitory allocation in universities, the assignment of tasks to workers, course assignment, kidney exchange, and so on. The most ancient and widely used random assignment rule is random serial dictatorship (henceforth RSD). Despite its transparency, \cite{bogomolnaia2001new} showed that RSD suffers from an efficiency loss in the sense of sd-efficiency, and they further proposed a new random assignment rule called the probabilistic serial rule (henceforth PS rule). The PS rule performs better in efficiency and is envy-free. \cite{bogomolnaia2001new}'s research had intrigued a large number of papers studying random assignment.  Recently, several authors  considered adaptations of  the immediate acceptance algorithm in random assignment problems.\footnote{Actually,  \cite{bogomolnaia2015random}, \cite{2018harless}, and \cite{2020Stepwise} have considered random assignment rules with similar intuition.} However, for the axiomatic characterization of  the immediate acceptance algorithm in random assignment problems, it is still an open question.
Our paper contributes in this direction.

The most natural approach to adapt immediate acceptance algorithm in random assignment environments is simply to use the random immediate acceptance (RIA) rule introduced in Section 3 of this paper. Unfortunately, this induces inefficiency. To avoid the inefficiency of the RIA rule illustrated in Example 1,  we propose a new rule called the \textbf{probabilistic rank rule (PR rule)}, the corresponding algorithm, we call it the \textbf{probabilistic rank algorithm}, is a generalization of the immediate acceptance algorithm. The PR rule first lets agents consume their favorite objects at equal speeds simultaneously.  Each  agent consumes her favorite object until the quota of the object is exhausted or the agent is satiated. And secondly, let agents consume their second-choice objects at equal speeds simultaneously.  Each  agent consumes her second-choice object until the quota of such object is exhausted or the agent is satiated,  and so on.

Next, we define two new axioms: sd-rank-fairness and equal-rank envy-freeness. ``Sd-rank-fairness'' means that whenever an agent gets a positive share of one object, then all objects that are preferred to this object by the agent have been fully allocated to some agents, and all agents who put this object in a higher preference ranks are satiated at this object. Sd-rank-fairness is a refinement of sd-efficiency. In deterministic setting, it amounts to favoring higher ranks  defined by \cite{kojima2014boston}, which is considered to be the main feature of the Boston mechansim. ``Equal-rank envy-freeness'' means that if two agents put an object at the same rank, then these two agents have no incentive to exchange their assignment for this object. Equal-rank envy-freeness is a refinement of equal treatment of equals.

Finally, we characterize the PR rule by sd-rank-fairness and equal-rank envy-freeness.

\subsection{Literature review}

The immediate acceptance rule is a widely used school choice rule around the world. \cite{abdulkadiroglu2003school} find various shortcomings of this rule and propose to substitute it with the student-optimal deferred acceptance rule and the top trading cycles rule. Subsequent studies further criticize the immediate acceptance for various reasons. Typical papers include \cite{ergin2006games} on the inefficiency of Nash equilibrium outcomes, \cite{chen2006school} on easy to manipulate in experiments, and \cite{pathak2008leveling} on ``uneven playing field''.
On the other hand, some recent research finds advantages of the immediate acceptance rule over the student-optimal deferred acceptance rule. Typical papers include \cite{abdulkadirouglu2011resolving}, \cite{miralles2009school}, and \cite{featherstone2016boston}. Considering the axiomatic approach, \cite{kojima2014boston} characterize the immediate acceptance rule and derive a priority profile from the axiomatic  properties of a rule, while \cite{afacan2013alternative} and \cite{chen2016new} characterize this rule on full priority domain.

In random environments, the probabilistic serial rule (PS rule) designed by \cite{bogomolnaia2001new} has been studied extensively. Typical papers include \cite{katta2006solution} on generalizing the PS rule to accommodate general preference domains, \cite{che2010asymptotic} on asymptotic equivalence of the PS rule and random serial dictatorship, and \cite{hugh2014experimental} on experimental study of the incentives of the PS rule. Considering the axiomatic approach, \cite{bogomolnaia2012probabilistic} and \cite{hashimoto2014two} characterize the PS rule, respectively. \cite{harless2019efficient} propose the ordered-claims-algorithm algorithm to construct sd-efficient probabilistic allocations of objects.

In addition to generalizing the immediate acceptance rule to random environments, our paper also proposes new axioms for random assignment rules. Parallel to our paper, \cite{featherstone2020rank} defines rank efficiency, which is a refinement of both sd-efficiency and ex post efficiency, and defines a class of linear-programming mechanisms with practical applications and satisfying rank efficiency. As to be shown in proposition 1, rank efficiency is independent with sd-rank-fairness.

\section{The model}

Let $N=\{1,\dotsc,n\}$ be a finite set of \textbf{agents} and $A=\{a_1,\cdots,a_m\}$ be a finite set of \textbf{object types}. For each $a\in A$, let $q_a$ denote the \textbf{supply} or quota of $a$. We assume $\sum_{a \in A }q_a\geq n$. Each agent $i\in N$ has a strict preference $P_i$ over ${A}$,
 namely a complete, transitive and antisymmetric binary relation over $ A$. Let  $ \mathscr{P}$ denote the set of all  all strict preferences on $A$.
 Introducing additional notation, for each $P_i \in \mathscr{P}$, let the corresponding weak preference of $P_i$ be $R_i$. That is, for all $P_i \in \mathscr{P}$
 and all $a_1,a_2 \in A$, $a_1 R_i a_2$ denotes either $a_1 P_i a_2$ or $a_1=a_2$.
  Let $P=(P_1,\cdots, P_n)\in \mathscr{P}^{n}$ be the preference profile. Given $a\in A$ and $P_i\in \mathscr{P}$, let $P_i(a)$ be the \textbf{rank} of object $a$ at $P_i$, i.e., if object $a$ is the $l^{th}$ choice of agent $i$ under $P_i$, then $P_i(a)=l$.
  Let $I(a,k)$ be the set of agents who put object $a$ at her  $k^{th}$ choice, that is, $I(a,k)=\{j : P_j(a)=k\}$.
   For each $i \in N$, $a\in A$ and $P_i \in \mathscr{P}$, denote by $U(P_i, a)=\{b\in A: b P_i a\}$ the \textbf{strict upper contour set} of $P_i$ at $a$, and $\widehat{U}(P_i, a)=U(P_i, a) \cup \{a\}$ the \textbf{weak upper contour set} of $P_i$ at $a$.

A \textbf{deterministic assignment} is represented by an $n\times m$ binary matrix $D$ such that for each $ i \in N$ and $a \in A$, $D_{ia}=0$ or $D_{ia}=1$, for each $ i \in N$, $\sum_{a \in A} D_{ia}= 1$, and for each $a \in A$, $\sum_{i \in N} D_{ia}\leq q_a$. Let $\mathscr{D}$ be the set of all possible deterministic assignments.
 For each  $D \in \mathscr{D}$ and $i \in N$,  the row vector $D_i$ denotes the assignment of agent $i$.
 For $D, D'  \in \mathscr{D}$, if $D_{ia}=D'_{ib}=1$ for $a, b \in A$ and $a P_i b$, then we also write $D_i P_i D'_i.$
 A deterministic assignment $D \in \mathscr{D}$ is \textbf{Pareto efficient} if there exists no $D' \in \mathscr{D}$ such that $D'_i R_i D_i$ for all $i\in N$, and $D'_i P_i D_i$ for some $i\in N$.
 A \textbf{deterministic rule} is a mapping $\varphi: \mathscr{P}^{n} \rightarrow \mathscr{D}$ which finds a deterministic assignment for each $P \in \mathscr{P}^{n}$.

A \textbf{random assignment} is represented by an $n\times m$ matrix $\pi$. Each element $\pi_{ia}$ in $\pi$ is a probability with which agent $i$ is assigned object $a$, such that for each $a \in A$, $\sum_{i \in N} \pi_{ia}\leq q_a$ , and for each $i \in N$, $\sum_{a \in A} \pi_{ia}= 1$. We assume that given a random assignment $\pi$, each row $\pi_i$ is the random allocation of agent $i$, i.e., $\pi_i$ is a probability distribution over $A$.
 Each column $\pi_a$ represents the probabilities shares of an object $a$ to agents. By the extension of Birkhoff-von Neumann theorem (\cite{birkhoff1946three}) of \cite{budish2013designing}, each such matrix can be represented by a convex combination of deterministic assignments. Let $\Pi$ be the set of all random assignments.
 For each agent $i \in N$ and $P_i, P'_i \in \mathscr{P}$, we say that

$( \romannumeral 1)$ $\pi_i$ \textbf{weakly stochastically dominates} $\pi'_i $, written as $\pi_i sd (P_i)\pi'_i$, if for each $a\in A$,

\[
\sum_{a' \in \widehat{U}(P_i,a)} \pi_{ia'} \geq \sum_{a' \in \widehat{U}(P_i,a)} \pi'_{ia'};
\]

$( \romannumeral 2)$ $\pi_i$ \textbf{stochastically dominates} $\pi'_i $, written as $\pi_i SD (P_i)\pi'_i$, if for each $a\in A$,

\[
\sum_{a' \in \widehat{U}(P_i,a)} \pi_{ia'} \geq \sum_{a' \in \widehat{U}(P_i,a)}\pi_{ia'}'\mbox {,} \mbox { and }\pi_i \neq \pi'_i.
\]

Throughout the paper, we fix $A,N$, and denote a \textbf{problem} by the preference profile $P$. A \textbf{rule} is a mapping $\varphi: \mathscr{P}^{n} \rightarrow \Pi$ which finds a random assignment for each problem. Given $P \in \mathscr{P}^{n}$, $i\in N$ and $a\in A$, let $P_{-i}$ be the preference profile of agents $N\backslash \{i\}$, $\varphi_i(P)$ be the random allocation of agent $i$ under $\varphi(P)$, and $\varphi_{ia}(P)$ be the assignment of  object $a$ to agent $i$  under $\varphi(P)$, respectively. When $\varphi$ is deterministic, for each $P \in \mathscr{P}^{n}$ and $a\in A$, we denote $\varphi_a(P)$ as the set of agents to whom object $a$ be assigned under $\varphi(P)$, and $\varphi_i(P)$ as the object that agent $i$ receives.

A rule $\varphi$ is \textbf{non-wasteful} if for each $P\in \mathscr{P}^{n}$, $a\in A$, and $i\in N$, $\varphi_{ia}(P)>0$ implies that $\sum_{j \in N }\varphi_{ja'}(P)=q_{a'}$ for all $a'\in U(P_i, a)$.

For random assignment problems, two efficiency notions are commonly used. A rule $\varphi$ is \textbf{ex post efficient} if for each $P \in \mathscr{P}^{n}$, $\varphi(P)$ can be decomposed as a convex combination over Pareto efficient deterministic assignments.
The following axiom is a stronger requirement. A rule $\varphi$ is \textbf{sd-efficient} if for each $P \in \mathscr{P}^{n}$, there exists no $ \pi \in \Pi$ such that $\pi_i sd(P_i) \varphi_i(P)$ for each $i\in N$ and $\pi \neq \varphi(P)$.

Recently, \cite{featherstone2020rank} proposes the axiom of rank efficiency. Specifically, define the \textbf{rank distribution} of an assignment $\pi$ to be
$N^{\pi}(k)=\sum_{i \in N} \sum _{a \in A}( \pi_{ia})_{P_i(a)\leq k}
$
for each $k\in \{1,\ldots, m\}$, $N^{\pi}(k)$ is the expected value of all agents getting their $k^{th}$ or better choice  under the assignment $\pi$. Given $P \in \mathscr{P}^{n}$, a random assignment $\pi$ is \textbf{rank-dominated} by another random assignment $\pi'$ at $P$ if the rank distribution of $\pi$ stochastically dominates that of $\pi'$, that is, $N^{\pi}(k) \geq  N^{\pi'}(k)$ for all $k$ and with the inequality strict for at least one $k$.
 A rule $\varphi$ is \textbf{rank efficient} if for each $P\in \mathscr{P}^{n}$, $\varphi(P)$ is not rank-dominated by any other random assignments.

For fairness notions, the following three axioms are used extensively. A rule $\varphi$ satisfies \textbf{equal treatment of equals} if for each $P \in \mathscr{P}^{n}$ and each pair of $i, j\in N$ such that $P_i=P_j$, $\varphi_i(P) = \varphi_j(P)$.
A rule $\varphi$ is \textbf{weakly sd-envy-free} if for each $P \in \mathscr{P}^{n}$, there exists no pair of $i, j\in N$ such  that $\varphi_j(P) SD(P_i) \varphi_i(P)$.
A rule $\varphi$ is \textbf{sd-envy-free} if for each $P \in \mathscr{P}^{n}$, and each pair of $i, j\in N$, $\varphi_i(P) sd(P_i) \varphi_j(P)$.

For incentive properties,  a rule $\varphi$ is \textbf{weakly strategy-proof} if for each $P \in \mathscr{P}^{n}$, there exists no $i \in N$ and $P_i' \in \mathscr{P}$ such that $\varphi_i(P_i', P_{-i}) SD({P}_i) \varphi_i( P)$.
A rule $\varphi$ is \textbf{strategy-proof} if for each $P \in \mathscr{P}^{n}$, there exists no $i \in N$ and $P_i' \in \mathscr{P}$ such that $\varphi_i(P_i', P_{-i}) sd({P}_i) \varphi_i( P)$.

\section{Inefficiency of random immediate acceptance rule}

Let $\mathscr{F}$ denote the set of all bijections from $\{1,2, \ldots, n\}$ to the set of agents $N$. We refer to each of these bijections as an order of agents. That is, for any $f\in \mathscr{F}$, agent $f(1)$ is first and agent $f(2)$ is second, and so on. We also refer to each $f\in \mathscr{F}$ as the priority order of agents. The smaller the number, the higher priority an agent has. That is, $f(1)$ has the highest priority and agent $f(2)$ has the second highest priority, and so on. Given an ordering of agents $f$, denote the \textbf{simple immediate acceptance rule} induced by $f$ by $\varphi^f$. For each preference profile $P$, $\varphi^f(P)$ is a deterministic assignment derived through the following steps:

\emph{Step $1$}, each agent applies to her first-choice object. If  the number of agents applying to an object, say $a$, is less than its quota $q_a$, then all the applicants are permanently accepted by object $a$. If the number of agents applying to object $a$ is greater than its quota, then object $a$ will accept agents with the highest $q_a$ ordering and reject the other applicants. The quota of one object reduces by the number of agents it accepts in this step.

\emph{Step $t(\geqslant2)$}, each agent who is rejected at Step $t-1$ applies to her $t^{th}$-choice object. If the number of agents applying to an object, say $a$, is less than its remaining quota, say $q_a^t$, then all the applicants are permanently accepted by object $a$. If the number of agents applying to object $a$ is greater than its remaining quota, then object $a$ will accept agents with the highest $q_a^t$ ordering and reject the other applicants. The quota of one object reduces by the number of agents it accepts in this step.

The algorithm terminates when each agent has been assigned an object.

The simple immediate acceptance rule based on any ordering $f$ does not treat agents symmetrically. For example, agent $f(1)$ always gets her top choice, but agent $f(n)$ gets whatever is remaining after every other agent has made their choices. To restore fairness, we consider the following lottery rule.

Let the \textbf{random immediate acceptance rule (RIA rule)} be denoted by $\varphi^{ria}$. For each preference profile $P$, $\varphi^{ria}(P)$ is defined as

\[\varphi^{ria}(P)=\sum_{f\in \mathscr{F}}\frac{1}{n!}\varphi^f(P) \]

\noindent That is, each simple immediate acceptance assignment is chosen with equal probability, or equivalently an ordering is randomly chosen with uniform distribution and the induced simple immediate acceptance rule is used. Although RIA has the desirable property of being easy to understand and implement, it has avoidable  and possibly severe inefficiency as the following example 1 shows.


\begin{example}\label{example:naive}
\textbf{Inefficiency with RIA.}  Let $N=\{1,2,3,4\}$ be the set of agents and $A=\{a,b,c,d\}$ be the set of objets.  Let agents' preferences be as specified in the table:

\begin{center}
\begin{tabular}{cccc}
$P_1$ & $P_2$ & $P_3$ & $P_4$ \\
\hline
$a$   & $a$   & $b$   & $b$   \\
$c$   & $d$   & $c$   & $d$   \\
$b$   & $b$   & $a$   & $a$   \\
$d$   & $c$   & $d$   & $c$
\end{tabular}
\end{center}

By the procedure of RIA, the final random assignment, denoted $\pi$,  is stochastically dominated by $\pi'$, and consequently is inefficient:
\[
\pi =
\begin{pmatrix}
1/2 & 0   & 3/8 & 1/8 \\
1/2 & 0   & 1/8 & 3/8 \\
0   & 1/2 & 3/8 & 1/8 \\
0   & 1/2 & 1/8 & 3/8
\end{pmatrix}
\text{ and }
\pi' =
\begin{pmatrix}
1/2 & 0   & 1/2 & 0   \\
1/2 & 0   & 0   & 1/2 \\
0   & 1/2 & 1/2 & 0   \\
0   & 1/2 & 0   & 1/2
\end{pmatrix}.
\]
\end{example}

\section{The probabilistic rank rule}

To avoid the inefficiency as illustrated in Example 1 above, we now introduce another generalization of immediate acceptance algorithm to the random environments and call this random assignment rule the \textbf{probabilistic rank (PR) rule}, denoted $\varphi^{pr}$. For each $P \in \mathscr{P}^{n}$, the PR assignment $\varphi^{pr}(P)$ is determined by the following \textbf{probabilistic rank algorithm}.

Firstly, each  agent consumes her favorite object until the quota of the object is exhausted or the agent is satiated. And secondly, let agents consume their second-choice objects at equal speeds simultaneously.  Each  agent consumes her second-choice object until the quota of such object is exhausted or the agent is satiated, and so on. To be specific,

\begin{itemize}
  \item Step $1$, only the first choices of the agents are considered. Each agent starts consuming their favorite objects {simultaneously} at equal speeds. Each  agent consumes her favorite object until the quota of the object is exhausted or the agent is satiated.  Remove the agents who are satiated at  their favorite objects.
  \item Step $k$, only the $k^{th}$ choices of the remaining agents are considered. Each remaining agent starts consuming her $k^{th}$ preferred object {simultaneously} at equal speeds. An agent  consumes her $k^{th}$ preferred object until the quota of the object is exhausted or the agent is satiated. Remove the agents who are satiated at  this step.
\end{itemize}

The algorithm ends when all agents have been assigned one unit of object in total.

\textbf{Remark:} In the final remark 6 of \cite{bogomolnaia2015random}, the author raised a rule she called the ``Boston mechanism''. The rule is essentially the same as the PR rule in our model setting. We believe that our rule is defined independently with \cite{bogomolnaia2015random}. Moreover, we characterize the PR rule, while \cite{bogomolnaia2015random} used this rule proving independence of axioms on the lexicographic preference domain.

The following example 2 illustrates the procedure of the PR algorithm.

\begin{example}
Let $N=\{1,2,3, 4\}$, $A=\{a,b,c, d\}$. Preferences are as follows:

\begin{center}
\begin{tabular}{cccccccc}
$P_1$ & $P_2$ & $P_3$ & $P_4$ && $P_1'$ & $P_1''$\\
\hline
$a$   & $a$   & $a$   & $b$  && $a$   & $a$  \\
$b$   & $c$   & $c$   & $a$  && $c$   & $c$  \\
$c$   & $d$   & $d$   & $c$  && $b$   & $d$  \\
$d$   & $b$   & $b$   & $d$  && $d$   & $b$
\end{tabular}
\end{center}

The PR rule finds the assignment through the following procedure. In the first step, agents $1,2,3$ start consuming object $a$ simultaneously at equal speeds, and each of them gets $1/3$ of object $a$, while agent $4$ starts eating object $b$ until she consumes $b$ away. Agent $4$ is then removed from the economy. In the second step, agent $1$ does not consume any object because the quota of object $b$ is zero now, while agents $2$ and $3$ consume object $c$ simultaneously  in the same speed until there is no object $c$ left, and each of them gets $1/2$ of object $c$. In the third step, agents $ 2,3$ start eating object $d$ simultaneously at equal speeds. Agents $2$ and $3$ stop when they are satiated at object $d$, and each of them gets $1/6$ of object $d$, while agent $1$ consumes nothing because object $c$ has been consumed way in the previous step. In the fourth step, agent $1$ consumes object $d$ until she is satiated at object $d$ and there is no object $d$ left. Agent $1$ gets $2/3$ of object $d$. The resulting PR assignment, $\varphi^{pr}(P)$, is given as follows:

\[
\varphi^{pr}(P)=
\begin{pmatrix}
1/3 & 0 & 0& 2/3 \\
1/3 & 0 & 1/2 & 1/6\\
1/3 & 0 & 1/2& 1/6 \\
0 & 1 & 0 & 0
\end{pmatrix}
\]

The PR rule is vulnerable to preference manipulation of agents. In other words, agents have strong incentives  to misreport their  preferences. Actually, $\varphi^{pr}$ is not weakly strategy-proof (and hence not strategy-proof). If agent $1$ reports different preferences like $P_1'$ or $P_1''$, then the PR assignment is given by:

\[
\varphi^{pr}(P_1', P_{2}, P_{3},P_4)=\varphi^{pr}(P_1'', P_{2}, P_{3},P_4)=
\begin{pmatrix}
 1/3 & 0 & 1/3 & 1/3 \\
 1/3 & 0 & 1/3 & 1/3\\
 1/3 & 0 & 1/3 & 1/3 \\
   0 & 1 & 0 & 0
\end{pmatrix}
\]
\noindent which is preferred (in the sense of stochastic dominance) by agent $1$ to $\varphi^{pr}(P)$. This shows that $\varphi^{pr}$ is not weakly strategy-proof.

Moreover, $\varphi^{pr}$ is not weakly sd-envy-free (and hence not sd-envy-free). Consider the random assignment $\varphi^{pr}(P)$. $\varphi^{pr}(P)$ is not weakly sd-envy-free because $\varphi^{pr}_2(P)$ and $\varphi^{pr}_3(P)$ stochastically dominates $\varphi^{pr}_1(P)$ for agent 1 at $P_1$.

$\varphi^{pr}$ is not rank-efficient either. Consider the problem and the random assignment $\varphi^{pr}(P)$ in example 1. It is easy to verify that the following assignment
\[
\pi=
\begin{pmatrix}
 1 & 0 & 0& 0 \\
 0 & 0 & 1/2 & 1/2\\
 0 & 0 & 1/2& 1/2 \\
 0 & 1 & 0 & 0
\end{pmatrix}
\]
rank dominates $\varphi^{pr}(P)$ at $P$.
\end{example}

\section{Two new axioms}

Although the PR rule violates some classical axioms for random assignment rules, it performs better if we use other criteria to evaluate it. We first introduce a new axiom: sd-rank-fairness. Sd-rank-fairness requires that if the assignment of an agent for an object is greater than zero, then all objects that are preferred to this object by the agent are fully allocated, and all agents who put the object in higher preference ranks get a surplus of one at this object, i.e., are satiated at this object. A rule satisfying sd-rank-fairness tries to assign an object to agents who put it in a higher preference rank, and only when assigning the object to agents with higher preference rank for this object is impossible, does it consider assigning it to agents who put the object in lower preference ranks.

\begin{Definition}
A rule $\varphi$ satisfies \textbf{sd-rank-fairness} if for each $P\in \mathscr{P}^{n}$ and each $i \in N$ and $a \in A$, $\varphi_{ia}(P)> 0$ implies that

$( \romannumeral 1)$ $\sum_{j \in N }\varphi_{ja'}(P)=q_{a'}$ for all $a'$ such that $a'P_i a$.

$( \romannumeral 2)$ $\sum_{a' \in \widehat{U}(P_j,a)} \varphi_{ja'}(P)=1$ for each $ j$ such that $P_j(a)< P_i(a)$.
\end{Definition}

\textbf{Remark:} Part $( \romannumeral 1)$ of sd-rank-fairness corresponds to non-wastefulness defined in Section 2 of this paper. Part $( \romannumeral 2)$ of sd-rank-fairness is equivalent to respect for rank defined by \cite{2018harless} and stepwise ordinal efficiency defined by \cite{2020Stepwise}.

\cite{bogomolnaia2001new} showed that sd-efficiency is a refinement of ex post efficiency. \cite{featherstone2020rank} proposed a new efficiency notion: rank efficiency. Rank efficiency is a refinement of both sd-efficiency and ex post efficiency. While sd-rank-fairness and rank efficiency are \textbf{independent} concepts, it is also true that sd-rank-fairness is a refinement of both sd-efficiency and ex post efficiency.

\begin{Proposition}
If a rule $\varphi$ satisfies sd-rank-fairness, then $\varphi$ also satisfies sd-efficiency.
\end{Proposition}

\begin{proof}
Suppose that there exists $P\in \mathscr{P}^{n}$ such that $\varphi(P)$ satisfies sd-rank-fairness, but violates sd-efficiency. By lemma 3 of \cite{bogomolnaia2001new}, there exist $i_1, i_2, \ldots, i_k$ and $a_1, a_2, \ldots, a_k$ such that $a_2 P_{i_1} a_1$ and $\varphi_{i_1 a_1}(P)>0$; $a_3 P_{i_2} a_2$ and $\varphi_{i_2 a_2}(P)>0$; \ldots; $a_1 P_{i_k} a_k$ and $\varphi_{i_k a_k}(P)>0$. Let $K=\mbox{Max}_{t\in \{ 1,2, \ldots, k\}}P_{i_t}(a_t)$. Without loss of generality, suppose that $P_{i_1}(a_1)=K$. Consider agent $i_k$ for whom $a_1 P_{i_k} a_k$ and $\varphi_{i_k a_k}(P)>0$. Because $\varphi_{i_k a_k}(P)>0$, $\sum_{a \in \widehat{U}(P_{i_k}, a_1)} \varphi_{i_ka}(P)< 1$, i.e., agent $i_k$ is not satiated at object $a_1$. By the definition of $K$, $P_{i_k}(a_1)<P_{i_k}(a_k)\leq  P_{i_1}(a_1)=K$. Then, we have that $P_{i_k}(a_1)<P_{i_1}(a_1)$ and agent $i_k$ is not satiated at object $a_1$ but $\varphi_{i_1 a_1}(P)>0$, which contradicts sd-rank-fairness (part $( \romannumeral 2)$) of $P$.
\end{proof}

Now, we discuss about sd-rank-fairness in deterministic environments. For deterministic assignment problems,  \cite{kojima2014boston}
introduce an axiom called favoring higher ranks.  A deterministic rule $\varphi$ \textbf{favors higher ranks} if for each $P \in \mathscr{P}^{n}$, each $i\in N$ and each $a \in A$,
\[
a \,\,P_i \,\,\varphi_i(P) \Rightarrow  |\varphi_a(P)|=q_a \,\, \& \,\, P_j(a)\leq P_i(a), \mbox{ for each } j\in \varphi_a(P).
\]
Conditional on an assignment being deterministic, sd-rank-fairness is equivalent to favoring higher ranks and remains a refinement of both sd-efficiency and Pareto efficiency. Formally,

\begin{Proposition}
Let $D$ be a deterministic assignment. Then,

$( \romannumeral 1)$ $D$ is sd-rank-fair if and only if it favors higher ranks.

$( \romannumeral 2)$ $D$ is sd-rank-fair implies that it is sd-efficient; however, the converse need not hold.
\end{Proposition}

\begin{proof}
Part $( \romannumeral 1)$: 
In the deterministic environment, sd-rank-fairness reduces to the following form: a deterministic assignment $D$ satisfies sd-rank-fairness means that, whenever $D_i=a$, then $D_j \in \widehat{U}(P_j,a)$ for any $ j$ such that $P_j(a)<P_i(a)$,  and
$|D^{-1}(a')|=q_{a'}$ for all $a'$ such that $a'P_i a$. To obtain the property of favoring higher ranks of $
D$ we only need to show that $R_j(a') \leq R_i(a')$ for all $a'$ such that $a'P_i a$ and all $j \in {D^{-1}(a')}.$
We argue by contradiction. Suppose there exits some $\tilde{a} P_i a $ and $\tilde{j} \in N$ such that $D_{\tilde{j}}=\tilde{a}$ and $P_{\tilde{j}}(\tilde{a}) > P_{i}(\tilde{a})$.
Then the sd-rank-fairness property implies that $D_i \in \widehat{U}(P_i,\tilde{a})$, and consequently $D_i P_i \tilde{a}$.
This contradicts $\tilde{a} P_i a=D_i.$ It is shown that sd-rank-fairness implies favoring higher ranks.

For the inverse implication, we suppose a deterministic assignment $D$ favors higher ranks.
For $D_i=a$, if $a' P_i a $ then $|D^{-1}(a')|=q_{a'}$ and $P_j(a') \leq P_i(a')$ for all $j \in D^{-1}(a')$.
 To obtain the sd-rank-fairness of $
D$ we only need to show that $D_j \in \widehat{U}(P_j,a)$ for any $ j$ such that $P_j(a)< P_i(a)$.
We argue by contradiction. Suppose there exits some  $\tilde{j} \in N$ such that $P_{\tilde{j}}(a)< P_i(a)$ and $aP_{\tilde{j}}D_{\tilde{j}}$.
Then according to the property of favoring higher ranks of $D$, the condition $aP_{\tilde{j}}D_{\tilde{j}}$ implies that
$P_k(a) \leq P_{\tilde{j}}(a)$ for all $k\in D^{-1}(a)$,
 and by $D_i=a$ we obtain $P_i(a) \leq P_{\tilde{j}}(a)$.
This contradicts the assumption $P_{\tilde{j}}(a)< P_i(a)$. It is shown that favoring higher ranks implies sd-rank-fairness.

For part $( \romannumeral 2)$, Proposition 1 of \cite{kojima2014boston} shows  that favoring higher ranks implies ex post efficiency. Furthermore, for deterministic assignment environments, Proposition 8 of \cite{featherstone2020rank} shows that ex post efficiency is equivalent to sd-efficiency. Then combining part $( \romannumeral 1)$ we obtain that sd-rank-fairness implies sd-efficiency.

Consider the following example: $N=\{1,2,3\}$, $A=\{a,b,c\}$ and each object type has only one copy. Assume that the preferences of agents are given as follows: $aP_1 b P_1 c$, $bP_2 a P_2 c$, $bP_3 a P_3 c$. For assignment $D$ such that $D_1=c$, $D_2=a$, $D_3=b$, it is easy to show $D $ is ex post efficient, and therefore is sd-efficient. However, one can see that agent 1 prefers $a$ to her assignment, but $a$ is assigned to agent 2, who put object $a$ at a lower rank than the rank of $a$ given by agent 1. Then $D$ does not favor higher ranks, and therefore is not sd-rank-fair. This example shows that sd-efficiency does not necessarily implies sd-rank-fairness.
\end{proof}

Next, we define the second new axiom: equal-rank envy-freeness. An assignment satisfies equal-rank envy-freeness, if two agents put an object in the same preference rank, then exchanging the assignments of the two agents for this object cannot increase the surplus at the same object for either of them.

\begin{Definition}
A rule $\varphi$ satisfies \textbf{equal-rank envy-freeness} if for each $P\in \mathscr{P}^{n}$, each $a\in A$ and $i,j \in N$ such that $P_i(a)=P_j(a)$, $\mbox{min} (\sum_{a^{'} \in U(P_i, a)} \varphi_{ia^{'}}(P)+ \varphi_{ja}(P), 1) \leq \sum_{a^{'} \in \widehat{U}(P_i, a)}  \varphi_{ia^{'}}(P)$.
\end{Definition}

\begin{Proposition}
If a rule $\varphi$ satisfies equal-rank envy-freeness, then it also satisfies equal treatment of equals.
\end{Proposition}

\begin{proof}
For two agents $i,j \in N$, if $P_i=P_j$, by the exchangeability between $i$ and $j$, it is easy to show that
equal-rank envy-freeness implies equal treatment of equals.
\end{proof}

\section{Characterizing the PR rule}

In this section, we propose a characterization of the PR rule. \cite{bogomolnaia2001new} showed that no random assignment rule satisfies strategy-proofness, sd-efficiency, and equal treatment of equals at the same time. The following result tells us that if we strengthen sd-efficiency and equal treatment of equals, and abandon strategy-proofness, we can achieve another rule: the PR rule.

\begin{Theorem}
A rule $\varphi$ satisfies sd-rank-fairness and equal-rank envy-freeness if and only if $\varphi=\varphi^{pr}$.
\end{Theorem}

\begin{proof}
We can see from the definition of the PR rule that it satisfies sd-rank-fairness and equal-rank envy-freeness. We need only to show the only if part. We proceed by contradiction.

Suppose that there exists another rule $\varphi$ satisfying sd-rank-fairness and equal-rank envy-freeness, but is different from the PR rule. Given $P \in \mathscr{P}^{n}$, denote the random assignment $\varphi^{pr}(P)$ and $\varphi(P)$ as $\pi$ and $\delta$, respectively. Suppose that $\pi \neq \delta$.

\emph{\textbf{Step \emph{1}}}: Suppose that for some $i \in N$ and $a\in A$ such that $P_i(a)=1$, $\pi_{ia}\neq \delta_{ia}$. There are two subcases.

\emph{\textbf{Step \emph{1.1}}}, $\delta_{ia}<\pi_{ia}$. According to our model setting, $0\leq \delta_{ia}\leq 1$ and $0\leq \pi_{ia}\leq 1$. If $\delta_{ia}<\pi_{ia}$, we have $0\leq \delta_{ia}< 1$ and $0< \pi_{ia}\leq 1$. By equal-rank envy-freeness of $\delta$ and $\pi$, $\delta_{ja}\leq \delta_{ja}\leq \pi_{ja}$ for each $j\in I(a,1)$ (Note that $I(a,k)=\{j : P_j(a)=k\}$.). Therefore, $\sum_{j \in I(a,1)} \delta_{ja}< \sum_{j \in I(a,1)} \pi_{ja}\leq q_a$. Consequently, two cases may arise: $( \romannumeral 1)$ the remaining supply of object $a$, which is $q_a-\sum_{j \in I(a,1)} \delta_{ja}$ is unassigned; $( \romannumeral 2)$ there exists an agent $j \in N$ such that $P_j(a)> 1$ and $\delta_{ja}> 0$. Next let's consider these two cases.

\textbf{Case} $( \romannumeral 1)$: the remaining supply of object $a$, which is $q_a-\sum_{j \in I(a,1)} \delta_{ja}$ is unassigned. We have proved proved $0\leq \delta_{ia}< 1$. Since each agent should be assigned a total object of 1, the above inequality implies that we can find an object $b\in A$ such that $ \delta_{i,b}>0$ and $aR_i b$. By sd-rank-fairness (part $\romannumeral 1$) of $\delta$, $ \delta_{i,b}>0$ and $aR_i b$ imply $\sum_{j \in  N} \delta_{ja}=q_a$ and further $\sum_{j \in I(a,1)} \delta_{ja}=q_a$ since object $a$ is only allocated among those who put it in the highest highest preference rank. This contradicts the assumption that the remaining supply of object $a$, which is $q_a-\sum_{j \in I(a,1)} \delta_{ja}$ is is unassigned. Therefore, case $( \romannumeral 1)$ is impossible.

\textbf{Case} $( \romannumeral 2)$: there exists an agent $j \in N$ such that $P_j(a)> 1$ and $\delta_{ja}> 0$. Now, $\delta_{ia}< 1$, $P_j(a)> 1$, and $\delta_{ja}> 0$ contradict sd-rank-fairness (part $( \romannumeral 2)$) of $\delta$. Therefore, case $( \romannumeral 2)$ is impossible.

\emph{\textbf{Step \emph{1.2}}}, $\delta_{ia}>\pi_{ia}$. This case is symmetric to the first one, therefore we omit the proof.

$\vdots$

\indent \emph{\textbf{Step} \textbf{\emph{k}}}: Suppose that for some $i \in N$ and $a\in A$ such that $P_i(a)=k$, $\pi_{ia}\neq \delta_{ia}$. There are two subcases.

\emph{\textbf{Step \emph{k.1}}}, $\delta_{ia}<\pi_{ia}$. Because $\sum_{a^{'} \in \widehat {U}(P_i, a)} \pi_{ia} \leq 1$, $\sum_{a' \in \widehat {U}(P_i, a)} \delta_{ia} < 1$. By equal-rank envy-freeness of $\delta$, $\delta_{ja}< \pi_{ja}$ for any $ j\in I(a,k)$. Let $\widehat{I}=\sum\limits_{l=1}^kI(a,l)$. Now we have, $\sum_{j \in \widehat{I}} \delta_{ja}< \sum_{j \in \widehat{I}} \pi_{ja} \leq q_a$. Consequently, two cases may arise: $( \romannumeral 3)$ the remaining supply of object $a$, which is $q_a-\sum_{j \in \widehat{I}} \delta_{ja}$ is unassigned; $( \romannumeral 4)$ there exists an agent $j \in N$ such that $P_j(a)> P_i(a)$ and $\delta_{ja}> 0$. Next let's consider these two cases one after another.

\textbf{Case} $( \romannumeral 3)$: the remaining supply of object $a$, which is $q_a-\sum_{j \in \widehat{I}} \delta_{ja}$ is unassigned. We have proved that $\sum_{a^{'} \in \widehat {U}(P_i, a)} \delta_{ia} < 1$. Since each agent should receive one copy of all objects in total. The above above inequality implies that we can find an object $b\in A$ such that $ \delta_{i,b}>0$ and $aP_i b$. By sd-rank-fairness (part $\romannumeral 1$) of $\delta$, $\sum_{j \in N} \delta_{ja}=q_a$ and further $\sum_{j \in \widehat{I}} \delta_{ja}=q_a$ since object $a$ is only allocated among those who put it in the highest $k$ preference rank. This contradicts the assumption that the remaining supply of object $a$, which is $q_a-\sum_{j \in \widehat{I}} \delta_{ja}$ is unassigned. Therefore, case $( \romannumeral 3)$ is impossible.

\textbf{Case} $( \romannumeral 4)$: there exists an agent $j \in N$ such that $P_j(a)> P_i(a)$ and $\delta_{ja}> 0$. Now, $\sum_{a^{'} \in \widehat {U}(P_i, a)} \delta_{ia} < 1$, $P_j(a)> P_i(a)$ and $\delta_{ja}> 0$ contradict sd-rank-fairness (part $( \romannumeral 2)$) of $\delta$.

\emph{\textbf{Step \emph{k.2}}}, $\delta_{ia}>\pi_{ia}$.  This case is symmetric to step $k.1$, therefore we omit the proof.
\end{proof}

\textbf{Independence of axioms: }First, we give the rule that violates only sd-rank-fairness. The uniform assignment rule due to \cite{chambers2004consistency} satisfies equal-rank envy-freeness, but violates sd-rank-fairness. Second, we give the rule that violates only equal-rank envy-freeness. The PR rule requires that in each step, each agent consumes the object simultaneously at an equal speed. When agents consume the objects at  different speeds, the new rule satisfies sd-rank-fairness but violates equal-rank envy-freeness.

\section{Final remarks}

Our analysis introduces and characterizes a new rule for probabilistic assignment, the PR rule. In contrast with the random immediate acceptance, the PR rule is sd-efficient and can be characterized by simple and intuitive axioms based on rank. In this paper we abstract from priorities in school choice problems like \cite{abdulkadirouglu2009strategy} and \cite{erdil2008s}. Nevertheless, the PR rule can be easily adapted to the setting of semi-strict priorities. Future works probably can further investigate the PR rule in the environment of  coarse priorities.

Finally, we compare the PR rule with random serial dictatorship (RSD), the probabilistic serial (PS) rule, and the random immediate acceptance (RIA). The following table 1 summarizes all axioms and the above three rules that we have discussed in this paper.

\begin{table}[H]
\caption{Random Assignment Rules and Axioms}
\begin{center}
\begin{tabular}{ccccc}
\hline
Axioms                    &  RSD & PS & RIA & PR \\
\hline
Weak Strategy-proofness   & $\surd$      & $\surd$     & $\times$  & $\times$\\
Strategy-proofness        & $\surd$      & $\times$    & $\times$  & $\times$\\
Equal treatment of equals & $\surd$      & $\surd$     & $\surd$   & $\surd$\\
Weak Sd-envy-freeness     & $\surd$      & $\surd$     & $\times$  & $\times$\\
Sd-envy-freeness          & $\times$     & $\surd$     & $\times$  & $\times$\\
Equal-rank envy-freeness  & $\times$     & $\times$    & $\times$  & $\surd$\\
Ex post efficiency        & $\surd$      & $\surd$     & $\surd$   & $\surd$\\
Sd-efficiency             & $\times$     & $\surd$     & $\times$   & $\surd$\\
Sd-rank-fairness          & $\times$     & $\times$    & $\times$  & $\surd$\\
Rank efficiency           & $\times$     & $\times$    & $\times$  & $\times$\\ \hline
\end{tabular}
\end{center}
\end{table}

\bibliographystyle{ecca}
\bibliography{bib}
\end{document}